\documentclass[
showpacs,preprintnumbers,amsmath,amssymb]{revtex4}
\usepackage{amsmath}
\usepackage{graphicx}
\usepackage{subfig}
\usepackage{hyperref}
\usepackage{amssymb}
\usepackage[english]{babel}
\usepackage{epsfig}
\usepackage{wasysym}
\usepackage{graphicx}
\usepackage{indentfirst}
\usepackage{amsmath}
\usepackage{amsfonts}
\usepackage{amssymb}
\usepackage{enumerate}
\hypersetup{colorlinks,citecolor=green,filecolor=magenta,linkcolor=red,urlcolor=cyan,pdftex}

\begin{document}


\title{Higher-Derivative  $f(R,\Box R, T)$  Theories of Gravity}

 \author{M. J. S. Houndjo$^{(a,b)}$\footnote{e-mail:
  sthoundjo@yahoo.fr}, M. E. Rodrigues$^{(c,d)}$\footnote{e-mail: esialg@gmail.com}, N.S. Mazhari$^{(e)}$\footnote{e-mail: najmemazhari86@gmail.com},  D. Momeni$^{(e)}$\footnote{e-mail: momeni-d@enu.kz} and R. Myrzakulov $^{(e)}$\footnote{e-mail: rmyrzakulov@gmail.com} }

  
\affiliation{$^a$ \, Institut de Math\'{e}matiques et de Sciences Physiques (IMSP), 01 BP 613,  Porto-Novo, B\'{e}nin\\
$^{b}$\, Facult\'e des Sciences et Techniques de Natitingou - Universit\'e de Parakou - B\'enin \\
$^{c}$ \, Faculdade de F\'isica, PPGF, Universidade Federal do Par\'a, 66075-110, Bel\'em, Par\'a, Brazil\\   
$^{d}$\,Faculdade de Ci\^encias Exatas e Tecnologia, Universidade Federal do Par\'a - Campus Universit\'ario 
de Abaetetuba, CEP 68440-000, Abaetetuba, Par\'a, Brazil\\
$^{e}$\, Eurasian International Center for Theoretical Physics and Department of
General \& Theoretical Physics, Eurasian National University, 
Astana 010008, Kazakhstan }  

\begin{abstract}
In literature there is a model of modified gravity in which the matter Lagrangian is coupled to the geometry via trace of the stress-energy momentum tensor $T=T_{\mu}^{\mu}$. This type of modified gravity is called as $f(R,T)$ in which $R$ is Ricci scalar $R=R_{\mu}^{\mu}$. We extend manifestly this model to include the higher derivative term $\Box R$. We derived equation of motion (EOM) for the model by starting from the basic variational principle. Later we investigate FLRW cosmology for our model. We show that de Sitter solution is unstable for a generic type of $f(R,\Box R, T)$  model. Furthermore we investigate an inflationary scenario based on this model. A graceful exit from inflation is guaranteed in this type of modified gravity.
\end{abstract}

\pacs{95.30.Sf; 98.80.-k;  04.60.-m}

\maketitle 

\section{Introduction}

Different types of the observational data, namely type Ia supernovae , cosmic microwave background (CMB) , large scale structure , baryon acoustic oscillations , and weak lensing indicate that we live in an accelerating epoch of the Universe \cite{Ri98,PeRa03} . A simple way to address this problem is to modify the original Einstein-Hilbert action by an arbitrary function of the curvature term(s) like $R,R_{\mu\nu},R^{\alpha}_{\beta\mu\nu},...$. This approach is called modified gravity and originally proposed in \cite{Buchdahl} and recently re-introduced to address current acceleration of the Universe appropriately 
 \cite{RevNoOd},\cite{Nojiri:2006ri}. If we replace the classical action of gravity by an arbitrary function of $R$, the Ricci scalar term, this type of modified gravity theories is called as 
$f(R)$ gravity \cite{Carroll:2003wy}:
\begin{eqnarray}
S=\frac{1}{2\kappa^2}\int d^4x\sqrt{-g}f(R)
\end{eqnarray}

 Because of simplicity and vest applications, several aspects of this type of models have been investigated in literature
\cite{viablemodels}. A significant observation is that  the $f(R)$ gravity is Lorentz invariant and it is independent from the structure of the Riemannian space which is under study \cite{Momeni:2015uwx}. We know that General relativity satisfies all solar tests by a highly precision so we expect that small deviations from this theory can satisfy these local tests. It is in great advantage that $f(R)$ gravity satisfies all local 
solar
system tests
\cite{solartests,Olmo07} and based on this fact, we are able to successfully reconstruct viable models of $f(R)$ gravity for cosmological applications
\cite{Hu:2007nk,solartests2,Sawicki:2007tf,Amendola:2007nt}.
$f(R)$ models are responsible for accelerating expansion can be used as matter content to describe rotation curves of different galaxies
without the need for dark
matter . This issue is vastly studied by authors 
\cite{Cap2,Borowiec:2006qr,Mar1,Boehmer:2007kx,Bohmer:2007fh}. Because of different types of $f(R)$ models there are several review works which they collected different aspects of these models(see for example
 \cite{RevNoOd,
SoFa08}.) \par
If we think about a weak violation of the equivalence principle, locally we can couple gravity to the matter part of any model. A simplest way is
coupling of an arbitrary function of the Ricci scalar $R$
with the matter Lagrangian density $L_ m$\cite{Bertolami:2007gv}. The model will be in the form $f(R,L_m)$ and has been investigated in different forms. For example the
case of the arbitrary couplings in both geometry and matter has been studied in
\cite{ha08}. Different types of such coupled models have been proposed and they opened a class of models as
 matter-geometry coupling models \cite{Bertolami:2007vu, ha10}. Not only metric approach in which the metric $g_{\mu\nu}$ is the variable of action but even
Palatini formulation of the non-minimal geometry-coupling models has been introduced and investigated by different researchers
 \cite{Pal}.  Also another version of these non-minimally coupled models has been presented 
 in \cite{Rlm},where they are
assuming that the gravitational Lagrangian is given by an
arbitrary function of the Ricci scalar $R$ and of the matter
Lagrangian $L_m$. Such non-minimally coupled models of gravity naturally appear in different areas, for example 
we can mention here a class of models in the form of  $f(R,L_m)$ gravity, this model originally
was proposed in \cite{Poplawski:2006ey}. In  \cite{Poplawski:2006ey}, it was argued that  $f(R,L_m)$ could represent a 
relativistic  covariant model of interacting dark energy. The cosmological constant $\Lambda$ in the
gravitational Lagrangian is defined by an arbitrary function of the trace of the associated 
energy-momentum  tensor $T_{\mu\nu}$. An alternative name for this model is 
``$\Lambda(T)$ gravity'' in which the cosmological constant is variable as a function of cosmological time . It was showed that the model has good agreement with recent cosmological
data and is consistent
with $\Lambda CDM$ paradigm, without the need to specify an exact
form of the function $\Lambda(T)$ \cite{Poplawski:2006ey}. We mention here that the model proposed in \cite{Poplawski:2006ey} has more generality than the Palatini $f(R)$
gravity, and reduces to the Palatini $f(R)$  when we neglect the pressure of
the matter.
\par
Among all different models for gravity in the class of non-minimally coupled paradigm, a simple but powerful model was proposed as 
the $f(R,T)$ modified
theories of gravity \cite{Harko:2011kv}. In this model, it was assumed that gravity was coupled to the matter through the trace of the energy(stress)-momentum tensor of matter $T=T_{\mu}^{\mu}$. This simple but efficient idea attracted several authors to investigate different aspects of the modern cosmology like exact solution of the cosmological evolution, black holes, thermodynamic laws and etc
  \cite{Momeni:2011am}-\cite{Baffou:2013dpa}
In the present work we generalized this type of models by taking into account the higher derivatives term $\Box{R}$. This term is actually appeared in string theory and so, it can be considered as a potentially important contribution to the original idea $f(R,T)$. It was proposed firstly that $f(R,\Box{R})$ in \cite{Hindawi:1995cu}. In that Ref.  \cite{Hindawi:1995cu}, the authors showed that these models reduced to scalar field models coupled to  gravity . Starting by a general higher-derivative gravity theory in action, one could show that after some conformal transformations , the model is equivalent to canonical Einstein gravity coupled to a finite number scalar fields, some numbers of which are propagating physically and  some of them are ghost-like and the  effective potential has a single, unstable stationary point \cite{Schmidt}, \cite{Wands}.

The present paper is organized as follows. In Sec. (\ref{section2}) we present our model and derive equations of the motion in metric formalism. In Sec. (\ref{reduction}) we investigate double-scalar field reduction of this type of modified gravity, in details. In Sec. (\ref{stability}) we study the local stability of the model in de Sitter background. In Sec. (\ref{geodesic}) we derive the modified geodesic equation. In Sec. (\ref{The Newtonian limit}) we study the weak field Newtonian limit and the problem of the precession of the perihelion of Mercury is investigated in Sec. (\ref{The precession of the perihelion of Mercury}). Sections (\ref{inflation}) and (\ref{Graceful exit from inflation }) are devoted to the inflationary epoch and the machinery of this type of modified gravity to explain early time acceleration. We summarize and conclude in Sec. (\ref{conclusion}).

\section{$f(R,\Box R, T)$ theory}\label{section2}
We start this work by writing the general gravitational action in the following form 
\begin{eqnarray}
S=\frac{1}{2\kappa^2}\int d^4x\sqrt{-g}f(R,T,\square R)+\int d^4x \sqrt{-g}\mathcal{L}_m\label{action}\,
\end{eqnarray}
where $R=R_{\mu\nu}g^{\mu\nu}$  is the curvature scalar, $T=T_{\mu\nu}g^{\mu\nu}$ the trace of the energy-momentum tensor and $\square=g^{\mu\nu}\nabla_{\mu}\nabla_{\nu}$ the d'Alembertian \footnote{We use the natural
system of units with $G=c=1$, so that the Einstein gravitational
constant is defined as $\kappa^2 =8\pi$.}
.\par 
By varying this action in metric formalism, one gets the following expression
\begin{eqnarray}
\delta S= \frac{1}{2\kappa^2}\int d^4x\left[f\delta\sqrt{-g}+\sqrt{-g}\left(f_{R}\delta R+f_{T}\delta T +f_{\square R} \delta \square R\right)+2\kappa^2 \delta \left(\sqrt{-g}\mathcal{L}_m\right)\right]\label{vS}
\end{eqnarray}
some supplementary variations are given as following:
\begin{eqnarray}
\delta\sqrt{-g}&=&-\frac{1}{2}\sqrt{-g}g_{\mu\nu}\delta g^{\mu\nu}\,,\label{vg}\\
\delta \Gamma^{\lambda}_{\;\;\mu\nu}&=&
\frac{1}{2}g^{\lambda\alpha}\left[\nabla_{\nu}\delta g_{\mu\lambda}
+\nabla_{\mu}\delta g_{\nu\lambda}-
\nabla_{\lambda}\delta g_{\mu\nu}\right]\,,\label{vGamma}
\end{eqnarray}
 and the Palatini's contracted equation reads 
\begin{equation}
\delta R_{\mu\nu}=\nabla_{\alpha}\delta \Gamma^{\alpha}_{\;\;\mu\nu}-\nabla_{\nu}\delta 
\Gamma^{\alpha}_{\;\;\mu\alpha}\,.\label{Palatini}
\end{equation}
Making use of (\ref{Palatini}), one gets 
\begin{eqnarray}
\delta R&=&R_{\mu\nu}\delta g^{\mu\nu}+g^{\mu\nu}\delta R_{\mu\nu}\\
&=&R_{\mu\nu}\delta g^{\mu\nu}+g^{\mu\nu}\left[\nabla_{\alpha}\delta \Gamma^{\alpha}_{\;\;\mu\nu}-\nabla_{\nu}\delta 
\Gamma^{\alpha}_{\;\;\mu\alpha}\right].
\end{eqnarray}
In order to rewrite the above equation, we use (\ref{vGamma}) and the relation 
\begin{eqnarray} 
\delta (g_{\alpha\nu}g^{\mu\nu})=0\Rightarrow g^{\mu\nu}\delta g_{\alpha\nu}=-g_{\alpha\nu}\delta g^{\mu\nu}\label{dg},
\end{eqnarray}
getting  
\begin{eqnarray}
\delta R= R_{\mu\nu}\delta g^{\mu\nu}+
g_{\mu\nu}\square\delta g^{\mu\nu}-\nabla_{\mu}\nabla_{\nu}
\delta g^{\mu\nu}.\label{vR} 
\end{eqnarray}
Now, one can calculate 
\begin{eqnarray}
\delta T&=&\delta \left(T_{\mu\nu}g^{\mu\nu}\right)=
T_{\mu\nu}\delta g^{\mu\nu}+g^{\mu\nu} \left(\delta T_{\mu\nu}\right)\nonumber\\
&=&\left[T_{\mu\nu}\frac{\delta g^{\mu\nu}}{\delta g^{\alpha\beta}}+g^{\mu\nu}\frac{\delta T_{\mu\nu}}{\delta g^{\alpha\beta}}\right]\delta g^{\alpha\beta}\nonumber\\
&=&\left[T_{\mu\nu}+\Theta_{\mu\nu}
\right]\delta g^{\mu\nu}\,\label{vT}
\end{eqnarray}
where 
\begin{equation}
\Theta_{\mu\nu}=g^{\alpha\beta}
\frac{\delta T_{\alpha\beta}}{\delta g^{\mu\nu}}\label{Theta}.
\end{equation}
On the other hand one has
\begin{eqnarray}
\delta \square R&=&\delta\left[
g^{\mu\nu}\partial_{\mu}
\partial_{\nu}R-g^{\mu\nu}
\Gamma^{\lambda}_{\;\;\mu\nu}
\partial_{\lambda}R
\right]\nonumber\\
&=&\nabla_{\mu}\nabla _{\nu}R \delta g^{\mu\nu}+\square \delta R-g^{\mu\nu}
\partial _{\lambda}R\delta 
\Gamma^{\lambda}_{\;\;\mu\nu}\,,
\end{eqnarray}
 which, using (\ref{vGamma}), (\ref{dg}) and (\ref{vR}), becomes 
\begin{eqnarray}
\delta \square R&=& \nabla_{\mu}\nabla _{\nu}R \delta g^{\mu\nu}+\square \Big(R_{\mu\nu}\delta g^{\mu\nu}+g_{\mu\nu}\square \delta g^{\mu\nu}-\nabla_{\mu}\nabla_{\nu}\delta g^{\mu\nu}\Big)-g^{\mu\nu}\partial_{\lambda}R\frac{1}{2}g^{\lambda\alpha}\Big(\nabla_{\mu}\delta g_{\nu\alpha}+\nabla_{\nu}\delta g_{\mu\alpha}-\nabla_{\alpha}\delta g_{\mu\nu}\Big)\nonumber\\
&=&\nabla_{\mu}\nabla _{\nu}R \delta g^{\mu\nu}+\square R_{\mu\nu}\delta g^{\mu\nu}+R\square \delta g^{\mu\nu} +g_{\mu\nu}\square^2 \delta g^{\mu\nu}+2g^{\alpha\beta}
\nabla_{\alpha}R_{\mu\nu}\nabla_{\beta}\delta g^{\mu\nu}
-\square \nabla_{\mu}\nabla_{\nu}\delta g^{\mu\nu}\nonumber\\
&-&\frac{1}{2}\Big[\nabla_{\mu}\left(-g^{\mu\nu}g_{\nu\alpha}\delta g^{\lambda\alpha}\right)+\nabla_{\nu}\left(-g^{\mu\nu}g_{\mu\alpha}\delta g^{\lambda\alpha}\right)-\nabla_{\alpha}\left(-g^{\mu\nu}g_{\mu\nu}\delta g^{\lambda\alpha}\right)\Big]\nabla_{\lambda}R\nonumber\\
&=&\Big[\nabla_{\mu}\nabla_{\nu}R
+\square R_{\mu\nu}+\Big(R_{\mu\nu}
\square+g_{\mu\nu}\square^2-\square\nabla_{\mu}
\nabla_{\nu}-\nabla_{\mu} R\nabla_{\nu}+2g^{\alpha\beta}\nabla_{\alpha}R_{\mu\nu}
\nabla_{\beta}\Big)\Big]
\delta g^{\mu\nu}\label{vsquareR}
\end{eqnarray}

With (\ref{vg}), (\ref{vR}), (\ref{vT}) and (\ref{vsquareR}), the variation of the action (\ref{vS}) becomes 
\begin{eqnarray}
&&\delta S=\frac{1}{2\kappa^2}\int d^4x \Big\{
-\frac{1}{2}\sqrt{-g}g_{\mu\nu}\delta g^{\mu\nu}f
+f_{R}\sqrt{-g}\Big(R_{\mu\nu}+g_{\mu\nu}\square 
-\nabla_{\mu}\nabla_{\nu}\Big)\delta g^{\mu\nu}
+\sqrt{-g}f_{\square R}\Big(\nabla_{\mu}\nabla_{\nu}R+\square R_{\mu\nu}+R_{\mu\nu}\square \nonumber\\
&&+g_{\mu\nu}\square^2-\square\nabla_{\mu}\nabla_{\nu}
-\nabla_{\mu}R\nabla_{\nu}+2g^{\alpha\beta}\nabla_{\alpha}R_{\mu\nu}
\nabla_{\beta}\Big)\delta g^{\mu\nu}+\sqrt{-g}\left(T_{\mu\nu}+\Theta_{\mu\nu}\right)f_{T}\delta g^{\mu\nu}+2\kappa^2\frac{\delta\left(\sqrt{-g}\mathcal{L}_m\right)}{\delta g^{\mu\nu}}\delta g^{\mu\nu}\Big\}\label{vS2}
\end{eqnarray}

We define the energy-momentum tensor from the matter Lagrangian $\mathcal{L}_m$ density by 
\begin{equation}
T_{\mu\nu}=-\frac{2}{\sqrt{-g}}\frac{\delta \left(
\sqrt{-g}\mathcal{L}_m\right)}{\delta g^{\mu\nu}}\,,
\end{equation}
and evidencing the term  $\sqrt{-g}\delta g^{\mu\nu}$ in  (\ref{vS2}), integrating by parts the third, the fourth , fifth\footnote{$\sqrt{-g}f_{\square R}R_{\mu\nu}\square \delta g^{\mu\nu}=\square \left(\sqrt{-g}f_{\square R}R_{\mu\nu} \delta g^{\mu\nu}\right)-\sqrt{-g}\delta g^{\mu\nu}\left(R_{\mu\nu}\square f_{\square R}+f_{\square R}\square R_{\mu\nu}\right)$}, sixth \footnote{$-\sqrt{-g}f_{\square R}\nabla_{\mu}R\nabla_{\nu}\delta g^{\mu\nu}=-\nabla_{\nu}\left(\sqrt{-g}f_{\square R}\nabla_{\mu}R\delta g^{\mu\nu}\right)+\sqrt{-g}\delta g^{\mu\nu}\left(\nabla_{\mu}R\nabla_{\nu}f_{\square R}+f_{\square R}\nabla_{\nu}\nabla_{\mu}R\right)$} seventh\footnote{$2\sqrt{-g}f_{\square R}g^{\alpha\beta}
\nabla_{\alpha}R_{\mu\nu}\nabla_{\beta}\delta g^{\mu\nu}=\nabla_{\beta}\left(2\sqrt{-g}f_{\square R}g^{\alpha\beta}
\nabla_{\alpha}R_{\mu\nu}\delta g^{\mu\nu}\right)-2\sqrt{-g}
\delta g^{\mu\nu}\nabla_{\beta}\left(f_{\square R}g^{\alpha\beta}\nabla_{\alpha}R_{\mu\nu}\right)$} terms of (\ref{vS2}), and considering the principle of  minima action ($\delta S=0$), we get the following equation of motion for the $f(R,\Box R, T)$ gravity:
\begin{eqnarray}\label{eom}
&&\left[f_{R}R_{\mu\nu}-\frac{1}{2}g_{\mu\nu}f+\left(g_{\mu\nu}\square-\nabla_{\mu}
\nabla_{\nu}\right)f_{R}\right]+\Big[2f_{\square R}\left(\nabla_{(\mu}\nabla_{\nu)}
R-\square R_{\mu\nu}\right)\nonumber\\
&&-\Big(R_{\mu\nu}\square
+g_{\mu\nu}\square^2-\square\nabla_{\mu}\nabla_{\nu}
-\nabla_{\mu}R\nabla_{\nu}+2g^{\alpha\beta}\nabla_{\alpha}
R_{\mu\nu}\nabla_{\beta}\Big)
f_{\square R}\Big]=\kappa^2 T_{\mu\nu}-f_{T}\left(
T_{\mu\nu}+\Theta_{\mu\nu}\right).\label{em}
\end{eqnarray}
As an unusual term, one performs $\Box R_{\mu\nu}$, considering that the covariant derivative of a covariant tensor of second order, leads to a covariant tensor of third order. Hence, one gets 
\begin{eqnarray}
&&\square R_{\alpha\beta}=g^{\mu\nu}\nabla_{\mu}
\nabla_{\nu}R_{\alpha\beta}=g^{\mu\nu}\nabla_{\mu}
\left[\partial_{\nu}R_{\alpha\beta}-
\Gamma^{\lambda}_{\;\;\nu\alpha}R_{\lambda\beta}
-\Gamma^{\lambda}_{\;\;\nu\beta}R_{\lambda\alpha}\right]
\nonumber\\
&&=g^{\mu\nu}\Big\{\partial_{\mu}\left[\partial_{\nu}R_{\alpha\beta}-
\Gamma^{\lambda}_{\;\;\nu\alpha}R_{\lambda\beta}
-\Gamma^{\lambda}_{\;\;\nu\beta}R_{\lambda\alpha}\right]-\Gamma^{\gamma}_{\;\;\mu\nu}\left[\partial_{\gamma}R_{\alpha\beta}-
\Gamma^{\lambda}_{\;\;\gamma\alpha}R_{\lambda\beta}
-\Gamma^{\lambda}_{\;\;\gamma\beta}R_{\lambda\alpha}\right]\nonumber\\
&&-\Gamma^{\gamma}_{\;\;\mu\alpha}\left[\partial_{\nu}R_{\gamma\beta}-
\Gamma^{\lambda}_{\;\;\nu\gamma}R_{\lambda\beta}
-\Gamma^{\lambda}_{\;\;\nu\beta}R_{\lambda\gamma}\right]-\Gamma^{\gamma}_{\;\;\mu\beta}\left[\partial_{\nu}R_{\alpha\gamma}-
\Gamma^{\lambda}_{\;\;\nu\alpha}R_{\lambda\gamma}
-\Gamma^{\lambda}_{\;\;\nu\gamma}R_{\lambda\alpha}\right]
\Big\}
\end{eqnarray}
It is remarkable that when $f_{\Box R}=0$,  the Eq. (\ref{em}) reduces to the Eq. in $f(R,T)$ gravity  \cite{Harko:2011kv}, also when $f_{\Box R}=f_{T}=0$, it is identified with the Eq.in $f(R)$ gravity \cite{Carroll:2003wy}.

\section{Scalar field reduction of $f(R,\Box R, T)$ theory}\label{reduction}
It is well-known that by starting with the Einstein-Hilbert action in a higher dimensional spacetime, and using the Kaluza-Klein reduction, we obtain a class of the Bergmann–Wagoner bi-scalar general action of scalar–tensor gravity \cite{Bamba:2014jua}. 
The same technique but with a slightly different approach could be used to reduce $f(R)$ gravity to the single scalar field model \cite{Carroll:2003wy}. In this Sec. we argue that our model given by (\ref{action}) can be reduced to a double scalar field model under scalar field reduction technique. Our motivation is inspired from the case $f_{T}=0$, was studied in  \cite{Hindawi:1995cu}. We use the same terminology of the Ref.  \cite{Hindawi:1995cu}.

Let us to consider an auxiliary Lagrangian 
\begin{eqnarray}
&& L=\lambda_1+\alpha\lambda_2^2+2f(T)+\mu(R-\lambda_1)+\mu_1(\Box R-\lambda_2).
\end{eqnarray}
Following the general procedure of scalar field reduction in modified gravity models  \cite{Hindawi:1995cu},the first step is to introduce a set of Lagrange multipliers $\mu,\mu_1$. As discussed in literature  \cite{Hindawi:1995cu}, we have introduced more than one  auxiliary fields. A systematic way is to eliminate $\lambda_2$
￼by solving its equation of motion, which reads $\mu_1=2\alpha\lambda_2$. From this equation it implies that $\lambda_2=\frac{\mu_1}{2\alpha}$. Using the 
back-substitution $\lambda_2$ into the action we obtain:
\begin{eqnarray}
&&S=\frac{1}{2\kappa^2}\int d^4x\sqrt{-g}\Big[\mu R+\mu_1\Box \lambda_1+\lambda_1-\mu\lambda_1-\frac{\mu_1^2}{4\alpha}+2f(T).
\Big]
\end{eqnarray}
Now, to put canonical kinetic energy for $\mu_1,\lambda_1$ , we define $\lambda_1=\chi_1+\psi_1,\mu_1=\chi_1-\psi_1$,so we obtain
\begin{eqnarray}
&&S=\frac{1}{2\kappa^2}\int d^4x\sqrt{-g}\Big[\mu R-(\nabla\chi_1)^2+(\nabla\psi_1)^2-(\mu-1)(\chi_1+\psi_1)-\frac{(\chi_1-\psi_1)^2}{4\alpha}+2f(T).
\Big]
\end{eqnarray}
Finally we make conformal transformation $\hat{g}_{\mu\nu}=e^{\chi}g_{\mu\nu}$, with $\chi=\log\mu$, the action transforms to the canonical form :
\begin{eqnarray}
&&S=\frac{1}{2\kappa^2}\int d^4x\sqrt{-\hat{g}}\Big[\hat{R}-\frac{3}{2}(\hat{\nabla}\chi)^2-e^{-\chi}(\hat{\nabla}\chi_1)^2+e^{-\chi}(\hat{\nabla}\psi_1)^2-e^{2\chi}\Big((e^{\chi}-1)(\chi_1+\psi_1)+\frac{(\chi_1-\psi_1)^2}{4\alpha}
\Big)
+2f(T).
\Big]
\end{eqnarray}
We proved that the 
 original higher-derivative gravity theory can be written in an  equivalent form in which the action is decomposed  to  Einstein-Hilbert action and  three scalar fields. We can show that in this equivalent form of action, one field is ghost-like. Furthermore, the interaction potential posses an unstable stationary (saddle)  point. This saddle point is located at  $\chi=\chi_1=\psi_1=0$. \par
Let us to consider a second simple example 
$f=\alpha+\beta R+\gamma R^2+\epsilon R\Box R+2f(T)$. The first two terms are Starobinsky's inflationary model \cite{Starobinsky:1980te}, and the next term is a non-trivial higher order correction. To reduce the model to the multi-scalar fields, let us to suppose that $\lambda_1=R$ and $\lambda_2=\Box R$ . We write the equation of motion for $\lambda_2$, we get
$\partial f/\partial\lambda_2=\epsilon\lambda_1$. The good point is this equation is  independent of $\lambda_2$. By introducing a pair of 
Lagrange multipliers we obtain:
\begin{eqnarray}
&&S=  \frac{1}{2\kappa^2} \int{ d^4x \sqrt{-g} \left[ 
         \alpha + \beta\lambda_1 + \gamma{\lambda_1}^2 
         + \epsilon\lambda_1\lambda_2 + \mu \left(R-\lambda_1\right) 
         + \mu_1 \left(\Box\lambda_1-\lambda_2\right) +2f(T)\right] }. 
\end{eqnarray}
As we observe, now we have some scalar fields. If we study the $\lambda_2$ equation of motion we read:
\begin{equation}
   \epsilon\lambda_1 = \mu_1,
\end{equation}
This equation cannot be solved with respect to the  $\lambda_2$ . However 
plugging this solution into the action , we can eliminate  both $\lambda_2$ and 
$\mu_1$,  the resulting action is 
\begin{eqnarray}
  && S = \frac{1}{2\kappa^2} \int{ d^4x \sqrt{-g} \left[ 
         \mu R - \epsilon\ (\nabla\lambda_1)^2 - \mu\lambda_1 
             + \alpha + \beta\lambda_1 + \gamma{\lambda_1}^2 
       +2f(T) \right] }.
\end{eqnarray}
Note that, here 
the kinetic energy for $\lambda_1$ is already in canonical form. Now we apply the
conformal transformation given by $e^\chi=\mu$ . Using this transformation, we can put the action in the 
completely canonical form
\begin{equation}
   S = \frac{1}{2\kappa^2} \int{ d^4x \sqrt{-\hat{g}} \left[ \hat{R }
         - \frac{3}{2} \left(\hat{\nabla}\chi\right)^2
         - \epsilon e^{-\chi} \left(\hat{\nabla} \lambda_1\right)^2 - V(\lambda_1,\chi) 
        +2f(T)  \right] }, 
\end{equation}
We see a potential term given by, 
\begin{equation}
   V(\lambda_1,\chi) = e^{-2\chi} \left( e^\chi\lambda_1 - \alpha -\beta\lambda_1 
                            - \gamma{\lambda_1}^2  \right)
\end{equation}
Our observations are as the following:
\begin{itemize}
\item  The original higher-derivative gravity is shown to be equivalent to 
ordinary Einstein gravity though now coupled to only two scalar fields. If we want to have non-ghost fields, we should make 
$\epsilon>0$, and $k=1$.
\item 
By investigating the potential for  function, we  conclude that 
$V(\lambda_1,\chi)$ has a single stationary point. It is an easy task to show that near this point, we have 
a stable minimum . This stable  point  corresponds to the 
anti-deSitter space with $\hat{R}=-2\alpha/\left(\beta^2-4\alpha\gamma\right)$ in the original higher-derivative theory. 

\item 
 When $k=1$ all of the scalar fields are non-ghost-like.

\end{itemize}

\section{Stability of de Sitter}\label{stability}
In cosmological backgrounds, in early or late time, de Sitter (dS) or  nearly dS is the dominant scenario,. An unstable dS is essential to exit inflation. In this section, we'll investigate stability of our modified gravity model under dS space. Let us to start by trace of (\ref{em}) is given by the following:
\begin{eqnarray}\label{trace}
&&f_{R}R-2f+3\square f_{R}+2f_{\square R}\square R-\Big(R\square
+3\square^2
-\nabla_{\mu}R\nabla^{\mu}+2g^{\alpha\beta}\nabla_{\alpha}
R\nabla_{\beta}\Big)
f_{\square R}=\kappa^2 T-f_{T}\left(
T+\Theta\right).\label{em}
\end{eqnarray}
Requiring $R=R_0$, we have de Sitter existence
condition in vacuum:
\begin{eqnarray}\label{ds1}
&&f_{R_0}R_0-2f-\Big(R_0\square
+3\square^2
\Big)
f_{\square R_0}=0.
\end{eqnarray}
Perturbing around dS space, namely $R=R_0+\delta R$,we attend at the
perturbation Eq.
\begin{eqnarray}&&
(Rf_{RR}-f_R)|_{0}\delta R+(Rf_{R\square R}-2f_{\square R}+3f_{RR})|_{0}\square(\delta R)+3f_{R\square R}\square^2(\delta R)=0\label{ds2}.
\end{eqnarray}
in which the scalar on effective mass and \emph{hyper -mass} parameters read as the following:
\begin{eqnarray}&&
M^2=\frac{(f_R-Rf_{RR})|_{0}}{(Rf_{R\square R}-2f_{\square R}+3f_{RR})|_{0}},\\&&
N^2=-\frac{3f_{R\square R}|_{0}}{(Rf_{R\square R}-2f_{\square R}+3f_{RR})|_{0}}
\end{eqnarray}
We should study linear stability of the Eq. (\ref{ds2}). It will be straightforward to define an appropriate function $\delta R =\Psi$, so we rewrite (\ref{ds2}) in the following form:
\begin{eqnarray}&&
-\square\Psi+N^2 \square^2\Psi+M^2\Psi=0\label{ds3}.
\end{eqnarray}
Thus if $N=0$ when  $f_{\square R}=0$,  we have unstable dS solution, in particular  case, when we have a pure  $f (R)$ model, we have the well known condition $\frac{f_R}{Rf_{RR}}|_{0}>1$. Generally, the characteristic Eq. (\ref{ds3}) has four distinct roots are given by following:
\begin{eqnarray}&&
\lambda=\pm\frac{\sqrt{2}}{2N}\sqrt{1\mp \sqrt{1-4(MN)^2}}\label{roots}.
\end{eqnarray}
It is impossible to keep all roots with $\mathcal{R}(\lambda_i)<0$, so dS is still unstable for a generic form of our model.

\section{The equation of motion of test particles  in $f(R,\Box R, T)$ gravity}\label{geodesic}
Taking into account the covariant divergence of Eq. (\ref{eom}), with the use of the following mathematical identity\cite{T. Koivisto}
\begin{eqnarray}
&&\nabla^{\mu}\left[f_{R}R_{\mu\nu}-\frac{1}{2}g_{\mu\nu}f+\left(g_{\mu\nu}\square-\nabla_{\mu}
\nabla_{\nu}\right)f_{R}\right]=0
\end{eqnarray} $f(R,\Box R, T)$ is an arbitrary function of the Ricci scalar $R$ , the trace of the stress-energy tensor $T$ and higher derivative term  $\Box R$, we obtain for the divergence of the stress-energy tensor $T_{\mu\nu}$ the equation
\begin{eqnarray}
&&\nabla^{\mu}T_{\mu\nu}=\frac{1}{\kappa^2+f_T}\Big[\nabla^{\mu}\Sigma_{\mu\nu}+(T_{\mu\nu}+\Theta_{\mu\nu})\nabla^{\mu}f_{T}-g_{\mu\nu}f_{T}\nabla^{
\mu}p \Big].
\end{eqnarray}
here $\Sigma_{\mu\nu}$ is defined by 
\begin{eqnarray}\label{Sigma}
&&\Sigma_{\mu\nu}\equiv  2f_{\square R}\left(\nabla_{(\mu}\nabla_{\nu)}
R-\square R_{\mu\nu}\right)-\Big(R_{\mu\nu}\square
+g_{\mu\nu}\square^2-\square\nabla_{\mu}\nabla_{\nu}
-\nabla_{\mu}R\nabla_{\nu}+2g^{\alpha\beta}\nabla_{\alpha}
R_{\mu\nu}\nabla_{\beta}\Big)
f_{\square R}
\end{eqnarray}
and furthermore 
\begin{eqnarray}
&&\nabla^{\mu}f_{T}=\partial^{\mu}Rf_{TR}+\partial^{\mu}f_{TT}+\partial^{\mu}\square R f_{T\square R}.
\end{eqnarray}

We argue here that in 
the general $f(R,\Box R, T)$  type gravity models,  the matter sector is not decoupled from the geometry part, as a consequence we conclude that the 
energy-momentum tensor of matter  sector is not covariantly conserved. We find that the test particles in this type of the modified gravity doesn't move 
on the geodesic lines. This situation is similar to the
 $f\left(R,L_m\right)$ models \cite{Rlm} and $f(R,T)$ gravity \cite{Harko:2011kv},
where the coupling between matter and geometry induces an
extra acceleration acting on the particle. In the present
Section, we derive the equation of motion of test particle in
$f(R,\Box R, T)$ gravity models.\par
Let us to suppose that the matter Lagrangian is given by 
 a perfect fluid with pressure $p$ and energy density $\rho$. In this case the
divergence of the stress-energy tensor is written in the following form,
\begin{eqnarray}\label{cons2}
&&\nabla^{\mu}T_{\mu\nu}=\frac{1}{\kappa^2+f_T}\Big[\nabla^{\mu}\Sigma_{\mu\nu}+(T_{\mu\nu}+\Theta_{\mu\nu})\nabla^{\mu}f_{T}-g_{\mu\nu}f_{T}\nabla^{
\mu}p \Big].
\end{eqnarray}
For projection formalism we need to introduce 
 the projection operator
$h_{\mu \lambda }=g_{\mu\lambda}-u_{\mu }u_{\lambda }$, obviously 
 we have $h_{\mu \lambda }u^{\mu}=0$ and
$h_{\mu \lambda }T^{\mu\nu}=-h_{\lambda }^{\nu}p$, respectively.

Explicitly, Eq.~(\ref{cons2}) can be written in the form
\begin{eqnarray}\label{precontr}
&&\nabla _{\nu }\left( \rho +p\right)u^{\mu }u^{\nu }+\left(
\rho +p\right) \left[ u^{\nu }\nabla _{\nu }u^{\mu }
+u^{\mu }\nabla _{\nu}u^{\nu}\right] -g^{\mu \nu }\nabla _{\nu }p
\nonumber\\
&&=-\frac{1}{8\pi +f_{T}\left( R,\Box R,T\right) }\left\{
T^{\mu \nu }\nabla _{\nu }f_{T}\left( R,\Box R,T\right)
+g^{\mu \nu }\nabla _{\nu}
\left[ f_{T}\left( R,\Box R, T\right) p\right] +\nabla_{\nu}\Sigma^{\mu\nu}\right\}\, .
\end{eqnarray}

By contracting Eq.~(\ref{precontr}) with $h_{\mu \lambda }$ we obtain

\begin{equation}
h_{\mu \lambda }u^{\nu }\nabla _{\nu }u^{\mu }
=\frac{\kappa^2h_{\lambda}^{\nu}\nabla_{\nu}p+h_{\mu\lambda}\nabla_{\nu}\Sigma^{\mu\nu}}{(\rho+p)(\kappa^2+f_{T})}
.
\end{equation}

After multiplying with $g^{\alpha \lambda }$ and by taking into account the famous
identity

\begin{equation}
u^{\nu }\nabla _{\nu }u^{\mu }=\frac{d^{2}x^{\mu }}{ds^{2}}
+\Gamma _{\nu\lambda}^{\mu}u^{\nu }u^{\lambda }\, ,
\end{equation}

we obtain the equation of motion of a test fluid in $f\left(R,\Box R,T\right)$
gravity as

\begin{equation}
\frac{d^{2}x^{\mu }}{ds^{2}}+\Gamma _{\nu \lambda }^{\mu }u^{\nu }u^{\lambda
}=f^{\mu }\, ,  \label{eqmot}
\end{equation}

where

\begin{equation}\label{fmu}
f^{\mu }=\frac{\kappa^2\left(g^{\mu \nu }-u^{\mu }u^{\nu }\right)\nabla _{\nu }p+h^{\mu}_{\beta}\nabla_{\nu}\Sigma^{\beta\nu}}{\left(\rho +p\right)\left[\kappa^2
+f_{T}\left(R,\Box R,T\right)\right]}\, .
\end{equation}

We can demonstrate that the new appeared extra-force $f^{\mu }$  is not perpendicular to the four-velocity,
$f^{\mu}u_{\mu }=0$ because the extra term $h^{\mu}_{\beta}\nabla_{\nu}\Sigma^{\beta\nu}$.
But when  $f_{\Box R}=0$, we re-obtain the equation of motion of $f(R,T)$ theory \cite{Harko:2011kv}. In this latter case, the  extra-force  will be perpendicular to the four-velocity.
The pressureless limit doesn't correspond to   a pressureless fluid (dust), consequently the motion of the test particles doesn't obey geodesic paths.

\section{The Newtonian limit}\label{The Newtonian limit}

The force term (\ref{fmu})  can be formally rewritten as the gradient of a super potential $W$ as follows,
\begin{equation}\label{Q}
\frac{\kappa^2\left(g^{\mu \nu }-u^{\mu }u^{\nu }\right)\nabla _{\nu }p+h^{\mu}_{\beta}\nabla_{\nu}\Sigma^{\beta\nu}}{\left(\rho +p\right)\left[\kappa^2
+f_{T}\left(R,T\right)\right]}=\nabla ^{\mu }\Big(\log \sqrt{W}\Big)\, ,
\end{equation}
It is easy task to derive the geodesic  Eq.~(\ref{eqmot})  from the following modified (actually fractional) action of a text point particle:
\begin{equation}
\delta S_{p}=\delta \int L_{p}ds=\delta \int \sqrt{W}\sqrt{g_{\mu \nu
}u^{\mu }u^{\nu }}ds=0\, ,  \label{actpart}
\end{equation}
here $S_{p}$ and $L_{p}=\sqrt{W}\sqrt{g_{\mu \nu }u^{\mu }u^{\nu
}}$ are the point like action and the point like Lagrangian for the test
particles, respectively. This modified action leads to the following form of the geodesic equation:
\begin{equation}
\frac{d^{2}x^{\mu }}{ds^{2}}+\Gamma _{\nu \lambda }^{\mu }u^{\nu }u^{\lambda
}+\left( u^{\mu }u^{\nu }-g^{\mu \nu }\right) \nabla _{\nu }\ln \sqrt{W}=0
\, .
\end{equation}
Note that in the GR limit, when $\sqrt{W}\rightarrow 1$ we find the standard geodesic motion.
For a barotropic fluid,
$p=w\rho ,w\ll 1$ we know that $\rho +p\approx \rho ,T=\rho
-3p\approx \rho$, respectively. Furthermore, we
assume that the function $f_T$ is a function of $T\approx \rho $
only. We  expand in series $f_T$ near a critical point $\rho_0$ :
\begin{eqnarray}
f_T\left(\rho \right)=f_T\left(\rho _0\right)
+\left(\rho -\rho _0\right)f_{TT}
|_{\rho=\rho_0} =8\pi \left[\alpha_0+ \beta_0\left(\rho -\rho
_0\right)\right]
\end{eqnarray}
here $\alpha_0=f_T\left(\rho _0\right)/8\pi $ and
$\beta_0=f_{TT}|_{\rho=\rho_0}/8\pi$. With this assumption we obtain:
\begin{equation}\label{Q2}
W\approx
{{\rm e}^{2\,\ln  \left( {\frac {C}{\beta_{{0}}}} \right) w \left( 1+
\alpha_{{0}}-\beta_{{0}}r_{{0}} \right) ^{-1}}}-2\,{{\rm e}^{2\,\ln 
 \left( {\frac {C}{\beta_{{0}}}} \right) w \left( 1+\alpha_{{0}}-\beta
_{{0}}r_{{0}} \right) ^{-1}}}w{\beta_{{0}}}^{-1}{\rho}^{-1}+O \left( {
\rho}^{-2} \right)
 ,
\end{equation}
We can show that the Eq.~(\ref{Q}) is valid in both the non-relativistic (Newtonian) and the extreme relativistic (Modified gravity and GR) regimes.


Now we can calculate  the Newtonian limit of our  modified gravity model. The weak field limit of the gravitational field is well described by the following metric in Newtonian gauge,
\begin{equation}
ds\approx \sqrt{1+2\phi -\vec{v}^{2}}dt
\approx \left( 1+\phi -\vec{v}^{2}/2\right) dt\, ,
\end{equation}

here $\phi $ is the Newtonian potential and $\vec{v}$ stands for the usual spaceline  velocity of the fluid.
We can approximate $\sqrt{W}\left(\rho \right)$ given by Eq.~(\ref{Q2}) as

\begin{eqnarray}
\sqrt{W}\approx 1+\frac{w}{\left(1+\alpha_0-\beta_0\rho _0\right)}
\ln \left[\frac{C\rho }{1+\alpha_0+\beta_0\left(\rho -\rho
_0\right)}\right]=1+U\left(\rho \right)\, ,
\end{eqnarray}

Here we defined an appropriate Newtonian potential $U(\rho)$. Now we should derive the weak field limit (first order) of the equations of motion of a test particle in this force field, it can be explored using 
the variational principle

\begin{equation}
\delta \int \left[ 1+U\left(\rho \right)
+\phi -\frac{\vec{v}^{2}}{2}\right] dt=0\, ,
\end{equation}

If we calculate this variational term, we obtain:

\begin{equation}
\vec{a}=-\nabla \phi -\nabla U\left(\rho
\right)=\vec{a}_{N}+\vec{a}_p+\vec{a}_{E}\, ,
\end{equation}

here $\vec{a}$ defines  the total  net non-relativistic acceleration of the system,and because the gravitational potential is assumed to be conserved, so we can relate the acceleration to the gradient of the Newtonian potential as follows
$\vec{a}_{N}=-\nabla \phi $. Now we can define  the Newtonian gravitational
acceleration by the following:

\begin{equation}
\label{fidr} \vec{a}_p
=-\frac{C}{1+\alpha_0-\beta_0\rho _0}\frac{1}{\rho }(\nabla p-\nabla \Sigma^{0x})=-\frac{1}{\rho
}(\nabla p-\nabla \Sigma^{0x}) ,
\end{equation}

 fixing the integration constant $C$ as
$C=1+\alpha_0-\beta_0\rho _0$, we obtain the final form of the acceleration term:

\begin{equation}
\vec{a}_{E}\left(\rho ,p\right)
=\frac{\beta_0}{1+\alpha_0-\beta_0\rho _0}\frac{\nabla p-\nabla \Sigma^{0x}}{1+\alpha_0+\beta_0\left(\rho -\rho
_0\right)}\, ,
\end{equation}
We mention here that thanks to the isotropy and homogeneity of FLRW metric, using (\ref{Sigma}) we observe that the higher order terms in $\Sigma^{0x}$ vanish, so this correction doesn't change the acceleration.
This extra acceleration  is induced form the  the modified gravity action.

\section{The precession of the perihelion of Mercury}\label{The precession of the perihelion of Mercury}
Solar system tests provide good test background for theories of gravity, because the parameters are estimated with a very high precision. In our model of gravity, the extra-force$f^{\mu}$ given in (\ref{fmu}) generated as an effect of coupling between matter sector and higher-derivative terms $\Box{R}$ from geometry. Its our chance that we are able to estimate this force term using  the orbital parameters of
the motion of the planets around the Sun. A standard method is to use the invariant properties  of the
 Laplace-Runge-Lenz vector  (LRL vector),is
defined as 
\begin{eqnarray}
\vec{A}=\vec{v}\times \vec{L}-\alpha \vec{e}_{r},
\end{eqnarray}
here by  $\vec{v}$ we mean the relative velocity vector from the planet with mass $m$ to the central Sun with mass $M_{\odot}$ and this vector is given by 
 $\vec{r}=r\vec{e}_{r}$ and the unit vector $\vec{e}_{r}=\frac{\vec{r}}{r}$ is the radial unit vector of the polar coordinate system $(r,\theta)$. As usual frame work for planet motion, we use a two-body picture in which the system (planet+Sun) moves with a relative momentum vector  $\vec{p}=\mu \vec{v}$ and the mass of this displacement vector is the
reduced mass and it is defined by  $\mu =mM_{\odot }/\left( m+M_{\odot}\right) $. The relative angular momentum $\vec{L}$ is defined in the standard form 
 $ \vec{L}=\vec{r}\times \vec{p}=\mu
r^{2}\dot{\theta}\vec{k}$, where $\vec{p}$ is the relative linear momentum of the reduced mass, and $\alpha
=GmM_{\odot}$ \cite{prec}. 
With gravitational field, the orbit is an elliptic with 
eccentricity
$e$, major semi-axis $a$, and period $T$. The equation of the
orbit is presented  by $\left( L^{2}/\mu\alpha \right) r^{-1}=1+e\cos
\theta $. The LRL vector can be redefined as
\begin{equation}
\vec{A}=\left( \frac{\vec{L}^{2}}{\mu r}-\alpha \right)
\vec{e}_{r}-\dot{r}L\vec{e}_{\theta }\, ,
\end{equation}
It is illustrative to see that the  
  derivative of $\vec{A}$ with respect to the polar
angle $\theta $ is related to the effective potential of the central force:
\begin{equation}
\frac{d\vec{A}}{d\theta }
=r^{2}\left[ \frac{dV(r)}{dr}-\frac{\alpha}{r^{2}}\right] 
\vec{e}_{\theta
}\, ,
\end{equation}
The potential term $V(r)$ consists of the Post-Newtonian potential,
$V_{PN}(r)=-\frac{\alpha}{r}-3\frac{\alpha ^{2}}{mr^{2}},\ \ \mu\approx m$, plus the additional relativistic contribution
from the general coupling between matter and geometry. This vector quantity is given by:
\begin{eqnarray}
&&\frac{d\vec{A}}{d\theta }=r^2\Big[6\frac{\alpha^2}{mr^3}+m\vec{a}_{E}(\vec{r})\Big]\vec{e}_{\theta}
\end{eqnarray}


 The change in
direction $\Delta \phi $ of the perihelion of the planet is given in terms of the a change of
$\theta $ from $0$ to  $2\pi $ , it is defined by the following term
\begin{eqnarray}
&&\Delta\phi =\frac{1}{\alpha e}\int_{0}^{2\pi }|\dot{\vec{L}}\times\frac{
d\vec{A}}{d\theta }| d\theta
\end{eqnarray} 
If we substitute the $\frac{d\vec{A}}{d\theta }$ and simplifying we obtain:

\begin{equation}\label{prec}
\Delta \phi =24\pi ^{3}\left( \frac{a}{T}\right) ^{2}\frac{1}{1-e^{2}}
+\frac{L}{8\pi ^{3}me}
\frac{\left( 1-e^{2}\right) ^{3/2}}{\left( a/T\right)^{3}}
\int_{0}^{2\pi }\frac{a_{E}\left[ L^{2}
\left( 1+e\cos \theta \right)^{-1}/m\alpha \right] }{\left( 1+e\cos \theta
\right)^{2}}
\cos \theta d\theta\, ,
\end{equation}%
here we use the identity  $\frac{\alpha}{L}=\frac{2\pi  a}{T
\sqrt{1-e^{2}}}$. As usual, the second term in (\ref{prec}) gives the contribution to the
perihelion precession through the coupling between
matter and higher-derivative terms in the geometry.

When the extra-force is
constant,  $a_E\approx$ constant, in the Newtonian limit the
extra-acceleration can be expressed in a similar form in $f(R,T)$ gravity
\cite{Bertolami:2007gv}.

Using the Eq.~(\ref{prec}) we estimate the perihelion precession:

\begin{equation}\label{prec1}
\Delta \phi =\frac{6\pi GM_{\odot}}{a\left( 1-e^{2}\right) }
+\frac{2\pi
a^{2}
\sqrt{1-e^{2}}}{GM_{\odot}}a_{E}\, ,
\end{equation}
here we substitute the Kepler's third law, $T^2=4\pi
^2a^3/GM_{\odot}$. For the sample planet as Mercury $a=57.91\times 10^{11}$
cm, and $e=0.205615$, respectively, while $M_{\odot }=1.989\times
10^{33}$ g, we estimate the difference 
$
\left(\Delta \phi
\right)_{E}=\left(\Delta \phi \right)_{obs}-\left( \Delta \phi
\right) _{GR}=0.17(\frac{arcsec}{century})
$
can be attributed to other
physical effects. Hence the observational constraints requires
that the value of the constant $a_E$
$a_E\leq 1.28\times 10^{-9} cm/s^2$.

\section{Inflationary dynamics}\label{inflation}
To explore the very early era of the whole Universe, the idea of inflation proposed 
~\cite{Inflation, N-I, Starobinsky:1980te}. Inflationary scenario, verified by several types of observational data by the recent cosmological observations 
such as the Wilkinson Microwave anisotropy probe (WMAP)~\cite{Komatsu:2010fb, Hinshaw:2012aka}, the Planck satellite~\cite{Planck:2015xua, Ade:2015lrj}, and the BICEP2 experiment~\cite{Ade:2014xna, Ade:2015tva}
on the quite tiny anisotropy of the cosmic microwave background (CMB) radiation. The first simple model was single scalar field inflation (inflaton) as new inflation~\cite{N-I}, with a different form of the scalar potential we have the chaotic inflation~\cite{Linde:1983gd}, 
natural inflation~\cite{Freese:1990rb}, and power-law inflation with the exponential inflaton potential~\cite{Yokoyama:1987an} and etc. Recently new models 
of single field inflaton 
have been proposed in Refs.~\cite{R-I-S, Bezrukov:2007ep},~\cite{Linde:1993cn},~\cite{H-I}. (for reviews, see, e.g.,~\cite{Lidsey:1995np, Lyth:1998xn, Gorbunov:2011zzc, MRV, Linde:2014nna}). 

Modified gravity realized inflation first time by the model proposed by 
Starobinsky ~\cite{Starobinsky:1980te, Vilenkin:1985md} . The model inspired from quantum corrections to the classical Einstein-Hilbert action 
 such as $R^2$ term. We know that 
 the Starobinsky or $R^2$ inflation in vacuum is equivalent to non-minimal Higgs inflation ~\cite{Bezrukov:2007ep}.
(for reviews
see, for instance,~\cite{Nojiri:2010wj, Nojiri:2008ku, Joyce:2014kja, Book-Capozziello-Faraoni, Capozziello:2011et, Koyama:2015vza, Bamba:2012cp, delaCruzDombriz:2012xy, Bamba:2015uma, Bamba:2013iga, Bamba:2014eea}). 
There are several extensions of the Starobisky model in other types of the modified gravity theories 
.~\cite{Inflation-M-G, LT}. Our aim in this section is to develop a consistent model of inflation in the framework of $f(R,\Box{R},T)$ gravity. \par
Let us to start by  assuming that the metric of the universe is the flat Friedmann-Lemaitre-Robertson-Walker (FLRW) one,
\begin{eqnarray}
ds^2=dt^2-a^2(t)\left(dx^2+dy^2+dz^2\right),\label{flatmetric}
\end{eqnarray}
and that the ordinary content of the universe is a perfect fluid, the energy-momentum tensor can be written as
\begin{eqnarray}
T_{\mu\nu}=\left(\rho+p\right)u_{\mu}u_{\nu}-pg_{\mu\nu}\,\,,
\end{eqnarray}
where $\rho$ and $p$ are ordinary energy density and the pressure. Moreover, we will consider that this fluid is a barotropic one such that the equation of state EoS is $p=\omega\rho$. Therefore, one can take the matter Lagrangian density as $\mathcal{L}_m=-\omega\rho$, such that  $\Theta_{\mu\nu}=-2T_{\mu\nu}-\omega\rho g_{\mu\nu}$. 
\par
Within the metric (\ref{flatmetric}) and the above expression of the tensor $\Theta_{\mu\nu}$, the generalized Friedmann equations read
\begin{eqnarray}
&&2Hf'''_{\Box R}-\left(2H^2+3\dot{H}\right)f''_{\Box R}-\left(5H^3+2H\dot{H}+\ddot{H}\right)f'_{\Box R}+2\left(-2H^2\dot{H}+6\dot{H}^2+3H\ddot{H}+\dddot{H}
\right)f_{\Box R}\nonumber\\&&
+Hf'_{R}+\left(H^2+\dot{H}\right)f_{R}-\frac{1}{6}f=\frac{\rho}{3}\left[\kappa^2+\left(1+\omega\right)f_T\right]\\&&
f''''_{\Box R}+5Hf'''_{\Box R}+\left(5\dot{H}-8H^2\right)f''_{\Box R}+\left(-23H^3+2H\dot{H}+4\ddot{H}\right)f'_{\Box R}\nonumber\\&&
+2\left(-2H^2\dot{H}+6\dot{H}^2+3H\ddot{H}+\dddot{H}
\right)f_{\Box R}-f''_R-2Hf'_R-\left(3H^2+\dot{H}\right)f_R+\frac{1}{2}f=\kappa^2\omega\rho
\end{eqnarray}
We introduce the e-folds number 
$N$ is used in the inflationary descriptions as
\begin{eqnarray}\label{N}
&&N=\log\Big[\frac{a(t_{end})}{a(t)}\Big],
\end{eqnarray}
By introducing $N$,  if we set $t_{end} $ as the ending time of inflation, after that $t>t_{end}$,  the universe enters into the  radiation-dominated era and the reheating processes is started and particle production will produce the structures. Therefore, the total e-folds of the inflationary era is given by:
\begin{eqnarray}
&&N_{inf}=N|_{t=t_{inf}}=\log\Big[\frac{a(t_{end})}{a(t_{inf})}\Big].
\end{eqnarray}

To have thermalization epoch, we suppose that $55< N < 65$. During inflation the Hubble parameter is almost a constant  as de Sitter space. We take the model as the following:
\begin{eqnarray}
&&f(R,\Box R,T)= R+\alpha R^2+\beta R \Box R.
\end{eqnarray}

 The slow-roll approximation parameters are given by the following expressions:
\begin{eqnarray}
&&\Box R\approx 3 H
 \dot{R}, \ \ 16 \alpha H^2\ll1,\ \ |2\beta H \dot{R}|\ll1.
\end{eqnarray}
The first FLRW equation will reduce to the following approximation form:
\begin{eqnarray}
&& H^2\approx \kappa^2\rho\Big(1+16\alpha H^2-2\beta H \dot{R}\Big)\label{eqinf}
\end{eqnarray}

We're looking for an inflationary solution as $H^2=H_{inf}^2(1+N)$, plugging this solution in (\ref{eqinf}) we obtain:
\begin{eqnarray}
&&\rho\approx \frac{H_{inf}^2(1+N)}{\kappa^2}\Big(1-16\alpha H_{inf}^2(1+N)^2-24\beta H_{inf}^4(1+N)^3\label{rho}.
\Big).
\end{eqnarray}
The dS solution is represented by $H_{dS}^2=H_{inf}^2(N_{inf}+1)$. A simple checking proves that $\epsilon\equiv \frac{H_{inf}^2}{2H_{dS}^2}\ll1$. 
Now we reconstruct the effective potential for a single non-interacting inflaton using (\ref{rho}). We assume that the single inflaton $\phi$ is decoupled from the gravitational part, so, the energy density is given by :
\begin{eqnarray}
&&\rho\approx \frac{\dot{\phi}^2}{2}+V(\phi)\approx V(\phi)
\end{eqnarray}
here we suppose that the scalar inflaton is  slowly rolling, so $\frac{\dot{\phi}^2}{2}\ll V(\phi)$. Using the expression of e folding (\ref{N}) we can write $\frac{d\rho}{d\phi}=-\frac{H}{\dot{\phi}}\frac{d\rho}{dN}$, and the Klein-Gordon equation is written in the following form:
\begin{eqnarray}
&&\frac{d\phi}{dN}\approx -\frac{1}{H(N)}\sqrt{\frac{d\rho/dN}{3}}\label{dphi}.
\end{eqnarray}
Now we substitue (\ref{rho}) in (\ref{dphi}) we obtain:
\begin{eqnarray}
&&\phi(N)\approx -\frac{1}{\sqrt{3}\kappa}\Big(\log(1+N)-16\alpha H_{inf}^2 (1+N)-9\beta H_{inf}^4(1+N)^4
\Big)
\end{eqnarray}
consequently we obtain the effective potential as follows:
\begin{eqnarray}
V(\phi) = \left\{ \begin{array}{lr}
{\frac {{H_{{\inf}}}^{2}{{\rm e}^{-\frac{1}{4}\,{\it LambertW} \left( -4\,B{
{\rm e}^{-4\,\kappa\,\sqrt {3}\phi}} \right) }}}{{{\rm e}^{\kappa\,
\sqrt {3}\phi}}{\kappa}^{2}}}-{\frac {{H_{{\inf}}}^{2}A{{\rm e}^{-\frac{3}{4}\,
{\it LambertW} \left( -4\,B{{\rm e}^{-4\,\kappa\,\sqrt {3}\phi}}
 \right) }}}{  {{\rm e}^{3\kappa\,\sqrt {3
}\phi}}  {\kappa}^{2}}}-\frac{8}{3}\,{\frac {{H_{{\inf}}}^{2}{
{\rm e}^{-\,{\it LambertW} \left( -4\,B{{\rm e}^{-4\,\kappa\,\sqrt 
{3}\phi}} \right) }}}{  {{\rm e}^{4
\kappa\,\sqrt {3}\phi}}  {\kappa}^{2}}}
\ ,  A\ll B \\
\frac{H_{inf}^2 \Big(1-A{{\rm e}^{-2\,{\it LambertW} \left( -A{{\rm e}^{-\kappa\,\sqrt {3}
\phi}} \right) -2\,\kappa\,\sqrt {3}\phi}}-\frac{8}{3}\,{{\rm e}^{-3\,{\it 
LambertW} \left( -A{{\rm e}^{-\kappa\,\sqrt {3}\phi}} \right) -3\,
\kappa\,\sqrt {3}\phi}}
\Big)
}{\kappa^2 {{\rm e}^{{\it LambertW} \left( -A{{\rm e}^{-\kappa\,\sqrt {3}\phi}}
 \right) +\kappa\,\sqrt {3}\phi}}},\ \ 
B \ll A
\end{array} \right. \ .
\end{eqnarray}
here $A\equiv 16\alpha H_{inf}^2, B\equiv 9\beta H_{inf}^4$ and the the ``LambertW" function satisfies 
\begin{equation}
 LambertW(x) e^{LambertW(x)} = x.
\end{equation}

\section{Graceful exit from inflation }\label{Graceful exit from inflation }
We need to exit the inflationary era because the thermal history will start after $t>t_{end}$ and we need also large scale structure formation. So, our scenario should  define a quasi (unstable) dS solution. In this section we 
 analyse the instability of the dS solution ($H = H_{inf} (> 0) $= constant) during inflation by taking the linear first perturbations of the Hubble parameter as follows \cite{Bamba:2015uxa}:
\begin{eqnarray}
H(t)\approx H_{inf}(1+\delta(t))
\end{eqnarray}
Where we suppose that $|\delta(t)|\ll1$, and thus $H_{inf}\delta(t)$  defines  the linear first order perturbation beyound the de Sitter solution $H_{inf}$. We write the first FLRW  as the following:
\begin{eqnarray}\label{hds}
&&-228\,H \beta\, \dot{H}\ddot{H} +24\,
\beta H \dot{H}\dddot{H} +\dot{H}+ H ^{2}+96\,\beta
\, H ^{4}\dot{H} +6\,\beta\, \dddot{H}\ddot{H} -12\,H
\alpha\ddot{H}
-84\,\alpha\, H ^{2}\dot{H}\\&&\nonumber-18\,H  \beta\ddddot{H}-108\, H ^{2}
\beta \dddot{H} -264\,\beta\,
H ^{2} \dot{H} ^{2}-108\,\beta\, H ^{3}\ddot{H} -12
\,\alpha\, \dot{H} ^{2}-24\,
H ^{4}\alpha-24\,\beta\,\dot{H} ^{3}=0
\end{eqnarray}
when $H=H_{dS}$, $(1-12 \alpha H_{dS}^2)=0$, $\alpha=\frac{1}{12 H_{dS}^2}$.  We perturb (\ref{hds}) , the associated differential equation for 
 $\delta(t)$ is forth order linear differential equation:
\begin{eqnarray}\label{delta}
&&H_{{
\inf}}(1-24\,{H_{{\inf}}}^{2}\alpha)+ \left( -96\,{H_{{
\inf}}}^{3}\alpha+2\,H_{{\inf}} \right) \delta+\left( 1-84\,\alpha\,{H_{{\inf}}}^{2}+96\,\beta\,{H_{{\inf}}}^{4}
 \right) \dot{\delta} \\&&\nonumber+\left( -12\,H_{{\inf}}\alpha-108\,\beta\,{H_{{
\inf}}}^{3} \right) \ddot{\delta}+ \left(  -108\,{H_{{\inf}}}^{2}\beta \right) \dddot{\delta
}-18\,H_{{\inf}}\beta\,\ddddot{\delta} =0.
\end{eqnarray}
An exponential function in the form $\delta(t)=e^{\lambda t}$ is supposed to be a solution of (\ref{delta}),where $\lambda$ is a constant, so that we can investigate the instability of the de Sitter solution. If we can find a positive solution of $\lambda$, the dS solution can be unstable. Therefore, the universe can exit from inflation . Eq. (\ref{delta}) reduces to a quartic equation  for $\lambda$, in the form
\begin{eqnarray}
&&a_4  \lambda^4+a_3\lambda^3+a_2\lambda^2+a_1\lambda+a_0=0. 
\end{eqnarray}

Using the  Vieta's formulas \cite{Vieta}:
\begin{eqnarray}
&&\sum  \lambda_i=108\,{H_{{\inf}}}^{2}\beta ,	
\\&&
\sum  \lambda_i \lambda_j=--\frac{1}{H_{inf}}-108\,\beta\,{H_{{
\inf}}}^{3}	,
\\&&
\sum  \lambda_i \lambda_j \lambda_k=8\,{H_{{
\inf}}}^{2}-2\,H_{{\inf}} 	
\\&&
\Pi  \lambda_i=-H_{{
\inf}}, 	
\end{eqnarray}
Because $\Pi  \lambda_i<0$, all roots could not have the same sign, it is possible to have $\lambda_1>,\lambda_{2,3,4}<0$, so the system could be unstable under linear perturbations. This feature is independent of the sign of $\beta$. This is because for $\lambda_1> 0$, the amplitude of $\delta(t)$ increases in time.  We note that even if the other forms of inflaton fields are supposed, the method to check the instability of the dS solution is basically the same as the one demonstrated above. We should examine the linear perturbations of the Hubble parameter by using the gravitational field equation in $f(R,\Box{R},T)$ with solutions for the equation of motions in terms of inflaton. Although we expect that , the form of the solution for the perturbations will be altered, but in principle we can find a solution to mimic the property that the dS solution is unstable.

\section{Conclusion}\label{conclusion}  
  If gravity is a fundamental force of nature, particular attention must be directed to the type, mechanism, arrangement of early and late time behavor of the model.  In spite of the modified theories  of gravity , this report merely presented and approved a matter-geometry coupling model of gravity, with higher derivative terms $\Box{R}$.  We developed systematically the model of gravity in the form of $f(R,T)$ to $f(R,\Box{R},T)$, a motivation is due to the scalar reduction of this types of models to bi-scalar models in which we have only one ghost field and furthermore, the vacuum is considered as a non-trivial state for theory. This theory for special cases, is free of ghost. It is possible to test it by solar system tests as well as inflationary data. Theory has a unstable de Sitter solution and it is possible to reconstruct families of scalar potentials for a prescribed form of the Hubble parameter as function of e-folding. It was demonstrated that in this theory, we can exit always from quasi de Sitter era to radiation domination epochs. So, we can consider it as a viable generalization of the Einstein-Hilbert action in favour of the modified theories of gravity.

\vspace{2cm}

{\bf Acknowledgement}: Manuel E. Rodrigues  
thanks UFPA, Edital 04/2014 PROPESP, and CNPq, Edital MCTI/CNPQ/Universal 14/2014,  for partial financial support.


\begin{thebibliography}{17}



\bibitem{Ri98}

S. Perlmutter et al. [SNCP Collaboration], Astrophys. J. 517, 565 (1999); A. G. Riess et al.[SNST Collaboration], Astron. J. 116, 1009 (1998); D. N. Spergel et al. [WMAP Collabora- tion], Astrophys. J. Suppl. 148, 175 (2003); ibid. 170, 377 (2007); E. Komatsu et al. [WMAP Collaboration], ibid. 180, 330 (2009); E. Komatsu et al. [WMAP Collaboration], Astrophys. J. Suppl. 192, 18 (2011); M. Tegmark et al., Phys. Rev. D 69, 103501 (2004); U. Seljak et al. [SDSS Collaboration], Phys. Rev. D 71, 103515 (2005); D. J. Eisenstein et al., Astrophys. J.
633, 560 (2005); B. Jain and A. Taylor, Phys. Rev. Lett. 91, 141302 (2003).

\bibitem{PeRa03}
P. J. E. Peebles and B. Ratra, Rev. Mod. Phys. \textbf{75}, 559 (2003);
T. Padmanabhan, Phys. Repts. \textbf{380}, 235 (2003).
\bibitem{Buchdahl}
 H. A. Buchdahl, Mon. Not. Roy. Astron. Soc. 150, 1 (1970).

\bibitem{RevNoOd}
  S.~Nojiri and S.~D.~Odintsov,
  Phys.\ Rept.\  {\bf 505}, 59 (2011)
  [arXiv:1011.0544 [gr-qc]].

\bibitem{Nojiri:2006ri}
  S.~Nojiri and S.~D.~Odintsov,
  eConf C {\bf 0602061} (2006) 06
   [Int.\ J.\ Geom.\ Meth.\ Mod.\ Phys.\  {\bf 4} (2007) 115]
  doi:10.1142/S0219887807001928
  [hep-th/0601213].

\bibitem{DeFelice:2010aj} 
  A.~De Felice and S.~Tsujikawa,
  Living Rev.\ Rel.\  {\bf 13}, 3 (2010)
  [arXiv:1002.4928 [gr-qc]].


\bibitem{Carroll:2003wy}
S. M. Carroll, V. Duvvuri, M. Trodden, and M. S. Turner,
Phys. Rev. D \textbf{70}, 043528 (2004).

\bibitem{viablemodels}
T.~Koivisto, Phys.\ Rev.\ D \textbf{76}, 043527 (2007);
A.~A.~Starobinsky, JETP Lett.\ \textbf{86}, 157 (2007);
B.~Li, J.~D.~Barrow, and D.~F.~Mota,
Phys.\ Rev.\ D \textbf{76}, 044027 (2007);
  A.~Azadi, D.~Momeni and M.~Nouri-Zonoz,
  Phys.\ Lett.\ B {\bf 670} (2008) 210,
  [arXiv:0810.4673 [gr-qc]];
S.~E.~Perez Bergliaffa, Phys.\ Lett.\ B \textbf{642}, 311 (2006);
V.~Faraoni, Phys.\ Rev.\ D \textbf{72}, 061501 (2005);
V.~Faraoni and S. Nadeau, Phys.\ Rev.\ D \textbf{72}, 124005 (2005);
A.~Abebe, D.~Momeni and R.~Myrzakulov,
  arXiv:1507.03265 [gr-qc];
 D.~Momeni, H.~Gholizade, M.~Raza and R.~Myrzakulov,
  Int.\ J.\ Mod.\ Phys.\ A {\bf 30} (2015) 16,  1550093,
  [arXiv:1502.05000 [gr-qc]];
L.~M.~Sokolowski, Class. Quantum Grav. \textbf{24}, 3391 (2007);
V.~Faraoni, Phys.\ Rev.\ D \textbf{75}, 067302 (2007);
M.~U.~Farooq, M.~Jamil, D.~Momeni and R.~Myrzakulov,
  Can.\ J.\ Phys.\  {\bf 91}, 703 (2013),
  [arXiv:1306.1637 [astro-ph.CO]];
M.~Jamil, F.~M.~Mahomed and D.~Momeni,
  Phys.\ Lett.\ B {\bf 702} (2011) 315,
  [arXiv:1105.2610 [physics.gen-ph]];
S.~Carloni, P.~K.~S.~Dunsby, and A.~Troisi,
Phys. Rev. D \textbf{77}, 024024 (2008);
S.~Nojiri and S.~D.~Odintsov, Phys.\ Lett.\ B \textbf{652}, 343 (2007);
S.~Tsujikawa, Phys.\ Rev.\ D \textbf{77}, 023507 (2008) ;
K.~N.~Ananda, S.~Carloni, and
P.~K.~S.~Dunsby, Phys. Rev. D \textbf{77}, 024033 (2008);
A. Guarnizo, L. Castaneda, and J. M. Tejeiro, arXiv:1002.0617v4 (2010).

\bibitem{Momeni:2015uwx} 
  D.~Momeni and R.~Myrzakulov,
  Astrophys.\ Space Sci.\  {\bf 360}, no. 1, 28 (2015),
  [arXiv:1511.01205 [physics.gen-ph]].

\bibitem{solartests}
T.~Chiba, Phys.\ Lett. B \textbf{575}, 1 (2003);
A.~L.~Erickcek, T.~L.~Smith, and M.~Kamionkowski, Phys.\ Rev.\ D
\textbf{74}, 121501 (2006);
T.~Chiba, T.~L.~Smith, and A.~L.~Erickcek, Phys.\ Rev.\ D
\textbf{75}, 124014 (2007);
S.~Nojiri and S.~D.~Odintsov, Phys. Lett. B \textbf{659}, 821 (2008);
S.~Capozziello, A.~Stabile, and A.~Troisi, Phys.
Rev. D \textbf{76}, 104019 (2007);
S.~Capozziello, A.~Stabile, and A.~Troisi,
Class. Quantum Grav. \textbf{25}, 085004 (2008).

\bibitem{Olmo07}
G.~J.~Olmo, Phys.\ Rev.\ D \textbf{75}, 023511 (2007).

\bibitem{Hu:2007nk}
W.~Hu and I.~Sawicki, Phys.\ Rev.\ D \textbf{76},
064004 (2007).

\bibitem{solartests2}
S.~Nojiri and S.~D.~Odintsov, Phys.\ Rev.\ D
\textbf{68}, 123512 (2003);Gen.\ Rel.\ Grav. 36 (2004) 1765;
V.~Faraoni, Phys.\ Rev.\ D \textbf{74}, 023529 (2006);
T.~Faulkner, M.~Tegmark, E.~F.~Bunn, and Y.~Mao, Phys.\ Rev.\ D \textbf{76},
063505 (2007); M.~Khurshudyan, N.~S.~Mazhari, D.~Momeni, R.~Myrzakulov and M.~Raza,
  Int.\ J.\ Theor.\ Phys.\  {\bf 54}, no. 2, 484 (2015)
  [arXiv:1403.0081 [gr-qc]];
C. S. J. Pun, Z. Kovacs, and T. Harko, Phys. Rev. \textbf{D 78}, 024043
(2008).

\bibitem{Sawicki:2007tf}
I.~Sawicki and W.~Hu, Phys.\ Rev.\ D \textbf{75},
127502 (2007).

\bibitem{Amendola:2007nt}
L.~Amendola and S.~Tsujikawa, Phys. Lett. B
\textbf{660}, 125 (2008).

\bibitem{odin}
S. Nojiri and S. D. Odintsov, Phys. Lett. B {\bf 657}, 238
(2007); S. Nojiri and S. D. Odintsov, Phys. Rev. D{\bf 77}, 026007
(2008).

\bibitem{Cap2}
S.~Capozziello, V.~F.~Cardone, and A.~Troisi, JCAP \textbf{%
0608}, 001 (2006);
S. Capozziello, V. F. Cardone, and A. Troisi, Mon. Not.
R. Astron. Soc. \textbf{375}, 1423 (2007).

\bibitem{Borowiec:2006qr}
A.~Borowiec, W.~Godlowski, and M.~Szydlowski,
Int. J. Geom. Meth. Mod. Phys. \textbf{4} (2007) 183

\bibitem{Mar1}
C.~F.~Martins and P.~Salucci, Mon. Not. R. Astron. Soc.
\textbf{381}, 1103 (2007).

\bibitem{Boehmer:2007kx}
C.~G.~Boehmer, T.~Harko, and F.~S.~N.~Lobo,
Astropart. Phys. \textbf{29}, 386 (2008).

\bibitem{Bohmer:2007fh}
C.~G.~Boehmer, T.~Harko, and F.~S.~N.~Lobo, JCAP
\textbf{03}, 024 (2008).

\bibitem{SoFa08}
T.~P.~Sotiriou and V.~Faraoni, Rev. Mod. Phys. {\bf 82},
451 (2010);
%
F.~S.~N.~Lobo,
arXiv:0807.1640 [gr-qc].
S. Capozziello and V. Faraoni, ``Beyond Einstein Gravity'', Springer, 2010.
%

\bibitem{Bertolami:2007gv}  O.~Bertolami, C.~G.~Boehmer, T.~Harko, and
F.~S.~N.~Lobo, Phys. Rev. D \textbf{75}, 104016 (2007).

\bibitem{ha08}
T.~Harko, Phys. Lett. B \textbf{669}, 376 (2008).

\bibitem{Bertolami:2007vu}  S. Nojiri and S. D. Odintsov, Phys. Lett. B {\bf
599}, 137 (2004);
V.~Faraoni, Phys. Rev. D \textbf{76}, 127501 (2007);
T.~P.~Sotiriou, Phys. Lett. B \textbf{664}, 225 (2008);
O.~Bertolami, T.~Harko, F.~S.~N. Lobo, and J.~Paramos, arXiv:0811.2876
(2008);
O. Bertolami and M. C. Sequeira,  Phys. Rev. D {\bf 79},
104010 (2009);
S. Nesseris, Phys. Rev. D {\bf 79}, 044015 (2009);
T. Harko, Phys. Rev. D {\bf 81}, 084050 (2010);
S. Thakur, A. A. Sen, and T. R. Seshadri, arXiv:1007.5250 (2010).

\bibitem{ha10}
T. Harko, Phys. Rev. D {\bf 81}, 044021 (2010).

\bibitem{Pal}
T. Harko, T. S. Koivisto and F. S. N. Lobo, arXiv:1007.4415
(2010).

\bibitem{Rlm}
T. Harko and F. S. N. Lobo, Eur. Phys. J. C {\bf 70}, 373
(2010).

\bibitem{Poplawski:2006ey}
N.~J.~Poplawski,
arXiv:gr-qc/0608031.






\bibitem{Harko:2011kv} 
  T.~Harko, F.~S.~N.~Lobo, S.~Nojiri and S.~D.~Odintsov,
  Phys.\ Rev.\ D {\bf 84}, 024020 (2011)
  doi:10.1103/PhysRevD.84.024020
  [arXiv:1104.2669 [gr-qc]].



\bibitem{Momeni:2011am} 
  M.~Jamil, D.~Momeni, M.~Raza and R.~Myrzakulov,
  Eur.\ Phys.\ J.\ C {\bf 72}, 1999 (2012),
  [arXiv:1107.5807 [physics.gen-ph]].

\bibitem{Alvarenga:2013syu} 
  F.~G.~Alvarenga, A.~de la Cruz-Dombriz, M.~J.~S.~Houndjo, M.~E.~Rodrigues and D.~Sáez-Gómez,
  Phys.\ Rev.\ D {\bf 87}, no. 10, 103526 (2013)
  [Phys.\ Rev.\ D {\bf 87}, no. 12, 129905 (2013)],
  [arXiv:1302.1866 [gr-qc]].



\bibitem{Alvarenga:2012bt} 
  F.~G.~Alvarenga, M.~J.~S.~Houndjo, A.~V.~Monwanou and J.~B.~C.~Orou,
  J.\ Mod.\ Phys.\  {\bf 4}, 130 (2013),
  [arXiv:1205.4678 [gr-qc]].


\bibitem{Jamil:2012pf} 
  M.~Jamil, D.~Momeni and R.~Myrzakulov,
  Chin.\ Phys.\ Lett.\  {\bf 29}, 109801 (2012),
  [arXiv:1209.2916 [physics.gen-ph]].


\bibitem{Sharif:2013gd} 
  M.~Sharif and M.~Zubair,
  J.\ Phys.\ Soc.\ Jap.\  {\bf 81}, 114005 (2012),
  [arXiv:1301.2251 [gr-qc]].


\bibitem{Momeni:2015gka} 
  D.~Momeni, R.~Myrzakulov and E.~Güdekli,
  Int.\ J.\ Geom.\ Meth.\ Mod.\ Phys.\  {\bf 12}, no. 10, 1550101 (2015),
  [arXiv:1502.00977 [gr-qc]].


\bibitem{Shabani:2013mya} 
  H.~Shabani and M.~Farhoudi,
  Phys.\ Rev.\ D {\bf 88}, 044048 (2013),
  [arXiv:1306.3164 [gr-qc]].



\bibitem{Sharif:2013yha} 
  M.~Sharif and M.~Zubair,
  J.\ Phys.\ Soc.\ Jap.\  {\bf 82}, no. 6, 064001 (2013),
  [arXiv:1310.1067 [gr-qc]].

\bibitem{Shamir:2015rva} 
  M.~F.~Shamir,
  Eur.\ Phys.\ J.\ C {\bf 75}, no. 8, 354 (2015),
  [arXiv:1507.08175 [physics.gen-ph]].

\bibitem{Baffou:2015dya} 
  E.~H.~Baffou, M.~J.~S.~Houndjo, M.~E.~Rodrigues, A.~V.~Kpadonou and J.~Tossa,
  arXiv:1509.06997 [gr-qc].

\bibitem{Sun:2015yga} 
  G.~Sun and Y.~C.~Huang,
  arXiv:1510.01061 [gr-qc].

\bibitem{Correa:2015qma} 
  R.~A.~C.~Correa and P.~H.~R.~S.~Moraes,
  arXiv:1509.00732 [hep-th].



\bibitem{Sharif:2014ioa} 
  M.~Sharif and M.~Zubair,
  Gen.\ Rel.\ Grav.\  {\bf 46}, 1723 (2014).

\bibitem{Shabani1:2014zza} 
  H.~Shabani and M.~Farhoudi,
  Phys.\ Rev.\ D {\bf 90}, no. 4, 044031 (2014),
  [arXiv:1407.6187 [gr-qc]].

\bibitem{Baffou:2013dpa} 
  E.~H.~Baffou, A.~V.~Kpadonou, M.~E.~Rodrigues, M.~J.~S.~Houndjo and J.~Tossa,
  Astrophys.\ Space Sci.\  {\bf 356}, no. 1, 173 (2015),
  [arXiv:1312.7311 [gr-qc]]


\bibitem{Hindawi:1995cu} 
  A.~Hindawi, B.~A.~Ovrut and D.~Waldram,
  Phys.\ Rev.\ D {\bf 53}, 5597 (1996),
  [hep-th/9509147].

\bibitem{Schmidt}
H.-J. Schmidt,, Class. Quantum Grav. 7 (1990), 1023–1031.



\bibitem{Wands}D. Wands, Class. Quantum Grav. 11 (1994), 269–279.
\bibitem{Bamba:2014jua} 
  K.~Bamba, D.~Momeni and R.~Myrzakulov,
  Int.\ J.\ Geom.\ Meth.\ Mod.\ Phys.\  {\bf 12}, no. 10, 1550106 (2015)
  doi:10.1142/S0219887815501066
  [arXiv:1404.4255 [hep-th]].
\bibitem{T. Koivisto}
T. Koivisto, Class. Quant. Grav. 23, 4289 (2006).


\bibitem{prec} B. M. Barker and R. F. O'Connell, Phys. Rev. D {\bf 10}, 1340
(1974); C. Duval, G. Gibbons, and P. Horvathy, Phys. Rev. D{\bf 43}, 3907
(1991).





\bibitem{Inflation} 
%
  K.~Sato,
  Mon.\ Not.\ Roy.\ Astron.\ Soc.\  {\bf 195}, 467 (1981);
%
  A.~H.~Guth,
  Phys.\ Rev.\ D {\bf 23}, 347 (1981).
%

\bibitem{N-I}
%
  A.~D.~Linde,
  Phys.\ Lett.\ B {\bf 108}, 389 (1982);
%
  A.~Albrecht and P.~J.~Steinhardt,
  Phys.\ Rev.\ Lett.\  {\bf 48}, 1220 (1982). 
%

\bibitem{Starobinsky:1980te} 
  A.~A.~Starobinsky,
  Phys.\ Lett.\ B {\bf 91}, 99 (1980).

\bibitem{Komatsu:2010fb}
E.~Komatsu {\it et al.} [WMAP Collaboration],
Astrophys.\ J.\ Suppl.\ {\bf 192}, 18 (2011) 
[arXiv:1001.4538 [astro-ph.CO]]. 

\bibitem{Hinshaw:2012aka} 
  G.~Hinshaw {\it et al.}  [WMAP Collaboration],
  Astrophys.\ J.\ Suppl.\  {\bf 208}, 19 (2013) 
  [arXiv:1212.5226 [astro-ph.CO]].

\bibitem{Planck:2015xua} 
  P.~A.~R.~Ade {\it et al.}  [Planck Collaboration],
  arXiv:1502.01589 [astro-ph.CO].

\bibitem{Ade:2015lrj} 
  P.~A.~R.~Ade {\it et al.}  [Planck Collaboration],
  arXiv:1502.02114 [astro-ph.CO].

%
\bibitem{Ade:2014xna}
  P.~A.~R.~Ade {\it et al.}  [BICEP2 Collaboration],
  Phys.\ Rev.\ Lett.\  {\bf 112}, 241101 (2014) 
  [arXiv:1403.3985 [astro-ph.CO]]. 

\bibitem{Ade:2015tva} 
  P.~A.~R.~Ade {\it et al.}  [BICEP2 and Planck Collaborations],
  Phys.\ Rev.\ Lett.\  {\bf 114}, 101301 (2015)
  [arXiv:1502.00612 [astro-ph.CO]].

\bibitem{Linde:1983gd} 
  A.~D.~Linde,
  Phys.\ Lett.\ B {\bf 129}, 177 (1983).

\bibitem{Freese:1990rb}
  K.~Freese, J.~A.~Frieman and A.~V.~Olinto,
  Phys.\ Rev.\ Lett.\  {\bf 65}, 3233 (1990).

\bibitem{Yokoyama:1987an}
  J.~Yokoyama and K.~I.~Maeda,
  Phys.\ Lett.\ B {\bf 207}, 31 (1988).


\bibitem{R-I-S}
%
  Y.~Hamada, H.~Kawai and K.~Y.~Oda,
  JHEP {\bf 1407}, 026 (2014)
  [arXiv:1404.6141 [hep-ph]];
  T.~Kobayashi and O.~Seto,
  Phys.\ Rev.\ D {\bf 89}, 103524 (2014)
  [arXiv:1403.5055 [astro-ph.CO]];
  M.~R.~Setare, D.~Momeni, V.~Kamali and R.~Myrzakulov,
  arXiv:1409.3200 [physics.gen-ph].
  Q.~Gao, Y.~Gong and T.~Li,
  Phys.\ Rev.\ D {\bf 91}, 063509 (2015)
  [arXiv:1405.6451 [gr-qc]];
  M.~Jamil, D.~Momeni and R.~Myrzakulov,
  Int.\ J.\ Theor.\ Phys.\  {\bf 54}, no. 4, 1098 (2015)
  doi:10.1007/s10773-014-2303-6
  [arXiv:1309.3269 [gr-qc]].
  M.~W.~Hossain, R.~Myrzakulov, M.~Sami and E.~N.~Saridakis,
  Phys.\ Lett.\ B {\bf 737}, 191 (2014)
  [arXiv:1405.7491 [gr-qc]];
  C.~Q.~Geng, M.~W.~Hossain, R.~Myrzakulov, M.~Sami and E.~N.~Saridakis,
  arXiv:1502.03597 [gr-qc];



%

\bibitem{Bezrukov:2007ep} 
  F.~L.~Bezrukov and M.~Shaposhnikov,
  Phys.\ Lett.\ B {\bf 659}, 703 (2008)
  [arXiv:0710.3755 [hep-th]].

\bibitem{Linde:1993cn} 
  A.~D.~Linde,
  Phys.\ Rev.\ D {\bf 49}, 748 (1994)
  [astro-ph/9307002].

\bibitem{H-I}
  T.~Kobayashi and O.~Seto,
  arXiv:1404.3102 [hep-ph];
  M.~Dine and L.~Stephenson-Haskins,
  arXiv:1408.0046 [hep-ph];
  J.~E.~Kim and D.~Y.~Mo,
  arXiv:1412.3544 [hep-ph];
  I.~Garc\'{i}a-Etxebarria, T.~W.~Grimm and I.~Valenzuela,
  arXiv:1412.5537 [hep-th];
  R.~Schimmrigk,
  arXiv:1412.8537 [hep-th];
S.~Nojiri, S.~D.~Odintsov, V.~K.~Oikonomou and E.~N.~Saridakis,
  arXiv:1503.08443 [gr-qc];
%

%

\bibitem{Lidsey:1995np}
  J.~E.~Lidsey, A.~R.~Liddle, E.~W.~Kolb, E.~J.~Copeland, T.~Barreiro and M.~Abney,
  Rev.\ Mod.\ Phys.\  {\bf 69}, 373 (1997)
  [astro-ph/9508078].

\bibitem{Lyth:1998xn}
  D.~H.~Lyth and A.~Riotto,
  Phys.\ Rept.\  {\bf 314}, 1 (1999)
  [hep-ph/9807278].

\bibitem{Gorbunov:2011zzc} 
D.~S.~Gorbunov and V.~A.~Rubakov, 
\textit{Introduction to the theory of the early universe: Cosmological perturbations and inflationary theory} 
(Hackensack, USA: World Scientific, 2011). 

\bibitem{MRV}
  J.~Martin, C.~Ringeval and V.~Vennin,
  Phys.\ Dark Univ.\  (2014)
  [arXiv:1303.3787 [astro-ph.CO]].

\bibitem{Linde:2014nna} 
  A.~Linde,
  arXiv:1402.0526 [hep-th].

\bibitem{Vilenkin:1985md} 
  A.~Vilenkin,
  Phys.\ Rev.\ D {\bf 32}, 2511 (1985).

%
\bibitem{Nojiri:2010wj}
S.~Nojiri and S.~D.~Odintsov,
Phys.\ Rept.\ {\bf 505}, 59 (2011)
[arXiv:1011.0544 [gr-qc]]. 

\bibitem{Nojiri:2008ku}
  S.~Nojiri and S.~D.~Odintsov,
  AIP Conf.\ Proc.\  {\bf 1115} (2009) 212
  [arXiv:0810.1557 [hep-th]].

\bibitem{Joyce:2014kja} 
  A.~Joyce, B.~Jain, J.~Khoury and M.~Trodden,
  Phys.\ Rept.\  {\bf 568}, 1 (2015)
  [arXiv:1407.0059 [astro-ph.CO]].

\bibitem{Book-Capozziello-Faraoni}
S.~Capozziello and V.~Faraoni,
\textit{Beyond Einstein Gravity}
(Springer, Dordrecht, 2010).

\bibitem{Capozziello:2011et}
S.~Capozziello and M.~De Laurentis,
Phys.\ Rept.\ {\bf 509}, 167 (2011)
[arXiv:1108.6266 [gr-qc]]. 

\bibitem{Koyama:2015vza} 
  K.~Koyama,
  arXiv:1504.04623 [astro-ph.CO].

\bibitem{Bamba:2012cp} 
  K.~Bamba, S.~Capozziello, S.~Nojiri and S.~D.~Odintsov,
  Astrophys.\ Space Sci.\  {\bf 342}, 155 (2012)
  [arXiv:1205.3421 [gr-qc]].

\bibitem{delaCruzDombriz:2012xy}
  A.~de la Cruz-Dombriz and D.~S\'{a}ez-G\'{o}mez,
  Entropy {\bf 14}, 1717 (2012)
  [arXiv:1207.2663 [gr-qc]].

\bibitem{Bamba:2015uma} 
  K.~Bamba and S.~D.~Odintsov,
  Symmetry {\bf 7}, 220 (2015)
  [arXiv:1503.00442 [hep-th]].

\bibitem{Bamba:2013iga} 
  K.~Bamba, S.~Nojiri and S.~D.~Odintsov,
  arXiv:1302.4831 [gr-qc].

\bibitem{Bamba:2014eea}
   K.~Bamba and S.~D.~Odintsov,
   arXiv:1402.7114 [hep-th]. 


\bibitem{Inflation-M-G}
  M.~Rinaldi, G.~Cognola, L.~Vanzo and S.~Zerbini,
  JCAP {\bf 1408}, 015 (2014)
  [arXiv:1406.1096 [gr-qc]];
  M.~Rinaldi, G.~Cognola, L.~Vanzo and S.~Zerbini,
  arXiv:1410.0631 [gr-qc];


\bibitem{LT}
  A.~B.~Lahanas and K.~Tamvakis,
  Phys.\ Rev.\ D {\bf 90}, 123530 (2014)
  [arXiv:1405.0828 [hep-th]].


\bibitem{Bamba:2015uxa} 
  K.~Bamba, S.~D.~Odintsov and P.~V.~Tretyakov,
  Eur.\ Phys.\ J.\ C {\bf 75}, no. 7, 344 (2015)
  [arXiv:1505.00854 [hep-th]].
\bibitem{Vieta}
 F. Viete,Opera mathematica. 1579. Reprinted Leiden, Netherlands, 1646.
  
\end{thebibliography}
\end{document}